\documentstyle[12pt]{article}
\newcommand{\be}{\begin{equation}}
\newcommand{\ee}{\end{equation}}
\newcommand{\bq}{\begin{eqnarray}}
\newcommand{\eq}{\end{eqnarray}}

\begin{document}

\pagestyle{empty}
\begin{flushright}
{\large UBCTP 92-26\\
 ITEP - M6/92 \\
July  1992}
\end{flushright}
\vspace{0.4cm}
\begin{center}
{\large \bf AREA LAW AND CONTINUUM LIMIT IN ``INDUCED QCD''
}\footnote{This work is supported in part by the Natural Sciences and
Engineering Research Council of Canada }\\ \vspace{1 cm} {\large I. I.
Kogan\footnote{Permanent Address: ITEP, 117295, Moscow, Russia.} A.
Morozov$^2$, G. W. Semenoff and N. Weiss}\\
\vspace{0.4 cm}
{ Department of Physics, University of British Columbia\\
Vancouver, B.C., Canada V6T1Z1}
 \end{center}
\vspace{0.4cm} \noindent
\begin{center}
{\bf Abstract}
\end{center}
We investigate a class of operators with
non-vanishing averages in a D-dimensional
matrix model recently proposed by Kazakov and Migdal. Among the
operators considered are ``filled Wilson loops" which are the most
reasonable counterparts of Wilson loops in the conventional Wilson
formulation of lattice QCD. The averages of interest are represented
as partition functions of certain 2-dimensional statistical systems
with  nearest neighbor interactions. The ``string tension"
$\alpha'$, which is the exponent in the area law for the ``filled
Wilson loop" is equal to the  free energy density
of the corresponding statistical system.
The continuum limit of the  Kazakov--Migdal model
corresponds to the critical point of this statistical
system.  We argue that in the large $N$ limit this critical
point occurs at zero temperature.
 In this case we express
$\alpha'$ in terms of the distribution density of
eigenvalues of the matrix-valued master field.
We show that the properties of the continuum limit and the description
of how this limit is approached
is very unusual and differs drastically  from  what occurs in both the
Wilson theory ($S\propto({\rm Tr}\prod U +{\rm c.c.})$) and in the
``adjoint'' theory ($S\propto\vert{\rm Tr}\prod U\vert^2$).
Instead, the continuum limit of the model appears to be intriguingly
similar to a $c>1$ string theory.

\newpage
\pagestyle{plain}
\setcounter{page}{1}
\section{Introduction}

In a recent paper, \cite{KM} V.Kazakov and A.Migdal  proposed  a new lattice
 gauge model  where the Yang-Mills interactions are induced by minimal
 coupling to a scalar field which transforms in the adjoint representation
 of the gauge group.  Their theory (to be referred below as the Kazakov-Migdal
model (KMM)) resembles a
D-dimensional matrix model  with partition function
of the form
 \footnote{ Throughout this paper we use square brackets to denote
 non-trivial measures of integration.
Thus, for example, $dX = \prod_{i,j} dX_{ij}$,
but the Haar measure $[dU]$ is
not simply $\prod_{i,j} dU_{ij}$, there is also a non-trivial factor,
see (\ref{(3'.2)}). In fact $[dU]$ will always denote the Haar measure
 over unitary matrices.
We shall denote normalized
averages in the theory (\ref{(1.1)}) by $\ll ~ \gg$,
while unnormalized integrals over unitary matrices with Haar measure $[dU]$
will be denoted by   $< ~ >$ .
Unitary matrices $U$ are taken to be elements of the group $SU(N)$.
If we were to allow $U$ to be in $U(N)$ then the $Z_N$-gauge
invariance of the KMM to be discussed throughout the text would extended to
a $U(1)$-gauge invariance. This does not influence our results, since
the model (\ref{(1.1)}) is  insensitive to the $U(1)$ factor in $U$.

Hermitian matrices such as  $\Phi$ and
$\Psi$ are {\it not} assumed to be traceless.
 However if the potential is quadratic the trace part of $\Phi(x)$
($\Phi(x) =
\phi_{{\rm tr}}(x)\cdot I + {\rm traceless~ part}$ )
 is easily separated from the traceless degrees of freedom
in (\ref{(1.1)}) giving rise to an extra factor
\bq
\prod_x \int d\phi_{\rm tr}(x) e^{N\left( -m^2\sum_x\phi_{\rm tr}^2(x) +
\sum_{<x,y>}\phi_{\rm tr}(x)\phi_{\rm tr}(y)\right) },
\nonumber
\eq
The critical value of the
 bare mass for this abelian  subsector of the KMM is
obviously $m_{\rm crit}^2 = D$. This is the the same
value of $m_{\rm crit}^2$ as is obtained for the non-
abelian components of $\Phi$ in the quasi--classical~ approximation. }
\bq
Z_{KMM}=\int [dU] d\Phi \exp\{-\sum_x {\rm tr}V(\Phi(x))
 +\nonumber
\eq
\bq\sum_{<x,y>}{\rm tr}~ \Phi (x) U(x,y)  \Phi (y)  U^{\dagger} (x,y)\}
\label{(1.1)}
\eq
Here $\Phi$ and $U$ are $N\times N$ Hermitean and unitary matrices
defined on the sites and the links of a D-dimensional rectangular lattice
respectively. Although the potential $V(\Phi)$ is quite general, for
the purpose of most of our work we can use the quadratic potential:
\bq
    V(\Phi) = m^2 \Phi^2
\label{(1.3)}
\eq
The effects of non-quadratic terms in the potential can be encoded in
quantities such as the
distribution function, $\rho(\phi)$,
of the eigenvalues of $\Phi(x)$.

For $D=4$ this model is a  lattice approximation to a Yang-Mills theory
alternate to the conventional Wilson lattice gauge theory.
Recall that the partition function of the Wilson theory is given by
\bq
\int [dU]   \exp \left(  \sum_{\Box} \frac{N}{g^2} W(\Box)  + c.c.\right),
\label{(1.2)}
\eq
where $W(\Box)$ is the elementary
Wilson loop, i.e. the trace of the product of link operators, around an
 elementary plaquette which we denote by $\Box$.

The main advantage of the KMM is the existence of two reformulations of
it, which, while being equivalent, look very different and emphasize
different features of the theory.

One reformulation arises from integration over the $\Phi$ variables in
(\ref{(1.1)}) and can be written as
\bq
Z_{KMM}=\int[dU] e^{-S_{eff}[U]} \propto\int [dU] \exp
\left(\frac{1}{2}\sum_\Gamma
\frac{1}{l(\Gamma)
m^{2l(\Gamma)}} \vert {W(\Gamma)}\vert^2 \right)~~~,
\label{(1.4)}
\eq
where the sum in the action is over {\it all}
 oriented contours $\Gamma$ on the
lattice and
\bq
    W(\Gamma) \equiv {\rm tr} \left( \prod_{<x,y>\in \Gamma} U(x,y)
\right);
\label{(1.5)}
\eq
with
\bq
    W(-\Gamma) = \overline{{W(\Gamma)}}; ~~~~~~~~
U(y,x)=U^{\dagger}(x,y).
\label{(1.6)}
\eq
In contrast to the Wilson theory, the effective action
depends on the modulus of $W(\Gamma)$, rather than on its
real part.  It also depends on contours of all
sizes rather than on just the elementary plaquette.

Another reformulation of the KMM arises by integrating first over $U$
rather than over $\Phi$. To do this, it is convenient
also to integrate out the angular
components of  $\Phi$ leaving an integral over
its eigenvalues
$~~\phi_1 (x)$,...,$\phi_N (x)~~$.
The partition function is then given by
\bq
Z_{KMM}\propto\int \prod_x \prod_i d{\phi_i (x)~
e^{-V(\phi_i(x))}~ \Delta^2 (\phi(x))} \prod_{<x,y>} I[\phi (x),
\phi (y)]
\label{(1.7)}
\eq
where
\bq
    \Delta(\phi) \equiv \prod_{i<j}^N (\phi_i - \phi_j),
\nonumber
\eq
and
\bq
    I[\phi,\psi] \equiv \int [dU] \exp \left(tr \Phi U
\Psi U^{\dagger}\right)
\label{(1.8)}
\eq

The first  representation (\ref{(1.4)}) has been used to give naive arguments
that
 the KMM (\ref{(1.1)}) is related to QCD (or, more precisely to quantum
{\it gluo}dynamics) if D=4. This idea, which we shall further
refer to as the ``naive continuum limit'', is that the action in
(\ref{(1.4)}) clearly has an absolute minimum at $U = I$.
Assuming that in the continuum limit the dominant contribution
to the partition function comes from the vicinity of this minimum
the effective action can  be computed by writing $U=\exp(iaA)$ (where $a$ is
the lattice
spacing). The result is derived in Ref. \cite{KM}:
\bq
    S\sim{\rm const} + {1\over g^2} \int d^4x {F_{\mu\nu}}^2 + ...
\label{(1.10)}
\eq
with
\bq
   \frac{ 1}{g^2} \sim N \int \frac{d^D p}{ p^{4}} \stackrel{D=4}
{\sim}~ N \log \frac{1}{a(m-m_{\rm crit})}.
\label{(1.11)}
\eq
We shall argue in this paper (see especially Section 6)
 that this argument is too naive and that the
approach to the continuum limit is much more complicated.

The second representation  (\ref{(1.7)}-\ref{(1.8)}) can be used for a much
more reliable investigation of the KMM. It may even be possible to solve the
KMM exactly for any N (although most of the progress has been made
for large $N$). The reason for these hopes is that the
integrals such as (\ref{(1.8)}) have  a pure algebraic interpretation so
that standard methods such as  Fourier analysis on co-adjoint
orbits or the Duistermaat-Heckmann (DH) integration formalism can be
applied to them.

 As argued in \cite{KM} the transformation from (\ref{(1.4)}) to
(\ref{(1.7)}) is highly non-trivial from a conceptual point of view and
is more or less equivalent to a summation of all planar diagrams in the
Yang-Mills theory.

In this paper we report some preliminary
considerations of correlation functions in  the KMM which should correspond
to the physical observables of ``induced QCD''. We shall concentrate our
 analysis on finite values of $N$. We shall then present arguments that
the features of the model for finite $N$  are sufficiently rich
and interesting that they may have  implications for the case of
infinite $N$.

   We shall see that the KMM  differs from the conventional Wilson
theory (\ref{(1.2)}) in many respects but  in particular in the properties
of the continuum limit. We show in Section 6 that the
probability distribution for the functional integral
near $U=I$ is much less sharp in the KMM than it
is in the Wilson theory. Moreover the continuum limit in the KMM appears
sensitive to the quantum measure in functional integral, not only to the
form of the action. This has a crucial effect on the normalization of
observable operators and probably on the other features of continuum
limit as well.

   Another purpose of this paper is to develop a formalism for evaluation of
the expectation values of operator observables
 in the KMM. We shall see that the most interesting
observables in the KMM are associated with
2-dimensional sublattices $S$ of the original $D$-dimensional
lattice. Vacuum expectation values of these quantities can be represented as
partition functions of $2d$ statistical systems defined on $S$.
We shall see that the Boltzmann factors on the links of this system are
fluctuating variables so that generically they behave in a manner similar
to spin glasses. In the zero-temperature limit, however,
they turn into ordered magnetic-type systems similar to $N$-state Potts
models. We shall argue that in the limit of large $N$ the zero-temperature
limit of the $2d$ system can be identified with the continuum limit of the KMM.
Furthermore we can show using these ideas that the area law for filled
Wilson Loops holds in this limit. The string
tension $\alpha'$ can be  expressed in terms of the density
 $\rho(\phi)$ of eigenvalues of  the
 master field $\Phi$. (See Sections 5 and 7 for details.)

Note that there is an ``adjoint'' model ``in between'' the KMM and Wilson
theory. Consider  a model with a partition function
\bq
Z_{\rm A}=\int [dU] \exp \sum_{\Box} \frac{1}{g^2} \vert
 {\rm tr} W(\Box)\vert^2
\label{(1'.3)}
\eq
Many of our comments in Sections 2 and 4 about the choice and the
 properties
of observables in the KMM are a result of
 $Z_N$ gauge invariance of the theory. They are thus equally true
for Eq. (\ref{(1'.3)}). The unusual behaviour of the continuum limit
which will be discussed in Sections 6 and 7 is peculiar to the KMM and is
drastically different from that in the theories ({\ref{(1.2)},\ref{(1'.3)}).
The reason for this difference is that in (\ref{(1.4)}) there is a sum
over {\it all} contours (not only over elementary plaquettes).
In any case one must keep firmly in mind that the KMM is significantly
different from both the
Wilson theory from the model of Eq. (\ref{(1'.3)}).

\section{Correlators and $Z_N$-invariance}

   Any lattice gauge theory possesses a gauge invariance with
 respect to
 transformations specified by
$P(x) \in U(N)$ at every site $x$ and  which act on the $\Phi$
and $U$-fields by
\bq
   \Phi(x) \rightarrow P(x)\Phi(x) P^\dagger (x) ,
\label{(2.1)}
\eq
\bq
  U(x,y) \rightarrow P(x)U(x,y)P^\dagger (y) .
\label{(2.2)}
\eq
The gauge invariant observables in lattice gluodynamics are made from
traces of products of $U$-matrices along closed oriented contours
$\Gamma$ on the lattice, i.e. conventional Wilson loops
(\ref{(1.5)},\ref{(1.6)}).

   The first important property of  the KMM $(1.1)$, which makes it
different from Wilson's theory, is that the space of gauge-invariant
operators must be further reduced \cite{KSW}. The reason for
 this is the presence
of an additional $Z(N)$ gauge symmetry: the freedom of independent
discrete transformations
\bq
    U(x,y) \rightarrow \omega(x,y) U(x,y);
\label{(2.3)}
\eq
\bq
    \omega(x,y)^N = 1 ,
\nonumber
\eq
so that  $\omega$ is any $N$-th root of unity. This
  additional gauge invariance excludes
the conventional Wilson loop $W(\Gamma)$ from the set of gauge invariant
observables (so that $\ll  W(\Gamma) \gg \equiv 0$). In fact
this symmetry excludes any product of Wilson loops as an observable
in the theory unless every link involved is passed an even number of
times, equally  in one and in the other direction.
To be precise, the difference between the number of $U$'s and
$U^{\dagger}$'s at every link should be divisible by $N$.
The set of gauge invariant observables in the
KMM thus consists of the products
\bq
    W(\Gamma_1, ..., \Gamma_k) \equiv \frac{1}{N} W(\Gamma_1)
 W(\Gamma
_2) ... W(\Gamma_k) ,
\label{(2.4)}
\eq
provided that
\bq
    \Gamma_1 + ... + \Gamma_k = 0~ [{\rm mod} N]
\label{(2.5)}
\eq
(The contours are, of course, taken to be oriented).

   The simplest examples of allowed operators are the
``Adjoint Wilson Loop'' (Fig.1)
\bq
    W(\Gamma, -\Gamma) = \frac{1}{N}\vert W(\Gamma)\vert^2,
\label{(2.6)}
\eq
the ``Baryon Loop''
\bq
    W_{\rm baryon}(\Gamma) = \frac{1}{N} W(N\Gamma)
\label{(2.6')}
\eq
(in which contour wraps $N$ times around $\Gamma$)
and the ``Filled Wilson Loop'' (FWL)  \cite{KSW},\cite{HS} (Fig.2)
\bq
\nonumber
   {\cal W}(S) \equiv &{1\over N} W(\Gamma,
-\Gamma_\alpha ~|~ \alpha \in {\rm plaquettes ~of~ S}) =
\eq
\bq
{1\over N} W(\Gamma)\prod_{ \rm plaquettes ~of~ S} \overline {W(\Box)}.
\label{(2.7)}
\eq
where $S$ is   2-dimensional surface on the
lattice with a given boundary $\Gamma = \partial S$. ${\cal W}$
involves the product of small Wilson loops over all the plaquettes
in $S$ as well as a big Wilson loop for $\Gamma$.

If this model has anything to do with QCD  it should posses
an operator like the Wilson loop which obeys an area law in the
confining phase.  (This property need not be related to existence
 of quark degrees of freedom). Neither the adjoint Wilson Loop nor
the Baryon Loop will suffice since they obey a perimeter rather
than an area law even in the standard Wilson theory for finite $N$.
The FWL (or (\ref{(2.11)}), to be exact) seems to be the only
reasonable counterpart of the conventional Wilson loop in the
framework of the KMM. There are several arguments
 to support this suggestion.

First of all FWL is a gauge-invariant operator (unlike
$W(\Gamma)$ which is not invariant under  local $Z(N)$). It
 can thus have a non-vanishing average value.
 Secondly, in the {\it naive} continuum limit the FWL reduces to
the conventional Wilson Loop
\bq
   {\cal W}(S) \sim {\rm tr}P\exp {i \oint_\Gamma {A_\mu (x) d x^ \mu}} .
\label{(2.8)}
\eq
(In particular, it depends only on $\Gamma = \partial S$.)  Indeed, if
 $U = \exp(iaA)$ then the plaquette loop

\bq
   W(\Box) = tr (I + ia^2F - \frac{1}{2}a^4 F^2 + ...) =
\nonumber
\eq
\bq
= N -{1\over2}a^4 F^2 + ... ,
\label{(2.9)}
\eq
where $F$ is the field strength. The last term vanishes in the naive
 continuum limit
and thus $W(\Box)$
tends to a field independent constant. (Note that we have used the fact
 that $tr F = 0$ in this derivation).
Therefore in this naive limit the plaquette loops do not
contribute at all. Their only role is to make the entire quantity
gauge invariant.  (Note that along with the ansatz $U = \exp (iaA)$ one
should consider a whole class of ansatze $U = \omega(x,y) \exp (iaA)$ with an
arbitrary $\omega(x,y) \in Z(N)$ at each link. This
would lead to the  cancellation of all contributions to $\ll W(\Gamma)\gg$,
but not to the average of the Filled Wilson Loop $\ll {\cal W}(S)\gg$.)
We shall further justify the use of the FWL by arguing later in the paper
that the expectation value  of the FWL obeys the area law,
\bq
    \ll  {\cal W}(S) \gg ~\sim \exp {\left( - \alpha' A(S) \right)},
\label{(2.10)}
\eq
where $A(S)$ stands for the area of $S$ (i.e. the number of elementary
plaquettes in $S$).

Before proceeding we must discuss the normalization of the FWL.
In the Wilson theory every loop $W(\Gamma)$ is usually
accompanied by a factor of $1/N$. According to this rule the FWL operator
should contain an additional factor $N^{-A(S)}$ in the definition
(\ref{(2.7)}).  This would  lead to cancellation of all the factors of $N$
in the naive continuum limit  leaving precisely a conventional Wilson
Loop with its correct normalization factor $1/N$.
Notice however that we have used a {\it different}
normalization factor in (\ref{(2.7)}) -- only a single factor of
 $\frac{1}{N}$.
The extra factor of $1/N^A$ makes a very big difference in that it can
lead to a spurious area law behaviour for the FWL.  We
shall argue below that this $1/N$ normalization is indeed the proper
 normalization. We
shall see that the naive continuum limit is unrelated to the true
continuum limit of the theory and that only this $1/N$ normalization leads
to finite averages in the  {\it true} continuum limit.

An additional complication which is introduced by the FWL  and
 is {\it unavoidable} in the KMM  is the dependence of its
expectation value on the surface $S$ rather than just on its
   boundary $\Gamma$.
It is thus nearly certain that the  correct
``physical'' operator (i.e. the operator which is to be compared
 with the Wilson loop of
continuum QCD) should contain an additional sum over all surfaces $S$
with the same boundary $\Gamma$:
\bq
   \hat {\cal W} (\Gamma) \equiv \int_{S, \partial S = \Gamma} [{\cal D}
S] ~{\cal W}(S)
\label{(2.11)}
\eq
with some (string-theory--like) measure $[{\cal D}S]$. This  is
important, for example, for the study of correlators of FWL operators
 (see Section $6$).
In this paper however we restrict ourselves to the problem of the
evaluation of $\ll {\cal W}(S) \gg$ for a {\it given} surface $S$.  (In
particular in order to derive the area law in the KMM it seems enough to
take for $S$ the surface of minimal area with the given boundary $\Gamma$.)

\section{Integrals over unitary matrices}

As discussed in the previous section we are interested in evaluating
the averages of operators such as  filled Wilson loops.
First note that averages of any operators which are invariant
under the $Z(N)$ gauge transformation (\ref{(2.2)}) are consistent with
the integration over angular variables which is implicit in the
transformation of (\ref{(1.1)}) into (\ref{(1.7)}). Thus the  the same
operators should be averaged in the ``Matrix Model'' approach
(\ref{(1.7)})
as are averaged in the original formulation (\ref{(1.1)}) of the KMM.
To evaluate
these averages one must first  study  integrals over
unitary matrices $U$ on a single link which are of the form
\bq
{I_{pq}[\phi ,\psi]}^{l_1k_1...l_qk_q}_{i_1j_1...i_pj_p} \equiv
{}~<U_{i_1j_1}...U_{i_pj_p}U^{\dagger k_1l_1}...U^{\dagger k_ql_q}>~
\equiv \nonumber
\eq
\bq
= \int [dU] \exp^{\sum_{i,j}\phi_i\psi_j \vert U_{ij}\vert^2}
 U_{i_1j_1}...U_{i_pj_p}U^{\dagger k_1l_1}...
U^{\dagger k_ql_q}.
\label{(3.1)}
\eq
Due to the
$Z(N)$ invariance under $U \rightarrow \omega U, ~~ U{^\dagger}
\rightarrow \omega^{-1} U^{\dagger}$ the integral is non-vanishing
only when the number of operators $U$ and $U^{\dagger}$ under
the integral differ by an integral multiple of $N$ i.e. $p-q = 0$ [mod $N$].
(This is the case for $U\in SU(N)$. If $U\in U(N)$ the condition is
that $p=q$).

The integrals (\ref{(3.1)}) are well known for $p$$=$$q$$=0$. In fact
there is an explicit formula
\bq
I_{00}[\phi ,\psi] =
\int  [dU] \exp(\sum_{i,j}\phi_i\psi_j \vert U_{ij}\vert^2)
 = {\rm const} \frac{\det_{ij} e^{\phi_i\psi_j}}
{\Delta(\phi)\Delta(\psi)}.
\label{(3.2)}
\eq
The standard references for this result are  \cite{Kh-Ch}.
There are, however, two more contemporary
 derivations which deserve mentioning.

One such derivation starts from an interpretation of $I_{00}$ as an
integral over the
co-adjoint orbit of $\Psi$:
\bq
I_{00}[\phi ,\psi] = {1\over{\Delta(\psi)}} \int [dX] e^{tr\Phi X}
\label{(3.3)}
\eq
where $X = U\Psi U^{\dagger}$ is a generic element of the orbit of $\Psi$
under the
action of $G/H$,
where $H$ is the Cartan subgroup of $G$ and $[dX]$ is the standard
symplectic measure
on this orbit. (Note  that the diagonal
components of $U$ do not act on a diagonal matrix $\Psi$.) This integral
was explicitly evaluated in \cite{AFS} with the use of the Gelfand-Tseytlin
parametrization of the orbit.

Another modern derivation of $I_{00}$ is based on application of the
Dui\-ster\-maat--Heckmann (DH) theorem \cite{DH}.
 The idea of this theorem is that
(under certain very restrictive conditions) an integral can be
substituted by a sum of quasi--classical contributions evaluated in the
vicinities of all the {\it extrema} (not only the minima) of the
action.  Applicability of the DH theorem to $I_{00}$ relies
on the fact that $I_{00}$ is essentially the integral over an orbit which
is a symplectic and symmetric manifold.

To see how the DH approach leads to the formula (\ref{(3.2)}) note that each
extrema of the action $\sum_{i,j} \phi_i \psi_j \vert U_{ij}\vert^2
$ as a function of $U$ is given by a product of any diagonal matrix
$U_d$ and any matrix of permutations $P$ i.e.  $U = U_d\times P$. Diagonal
matrices do not contribute to the action, and thus the sum in the DH
formula goes over all permutations of $N$ natural numbers $1...N$.  In
order to evaluate the contribution of any given extremum one needs to
find the determinant of quadratic fluctuations about this extremum. To do
 this we write
$U=P e^{iH}$ where $H$ is a Hermitean matrix. The action is then expanded
to quadratic order in $H$:
\bq
\sum_{i,j}\phi_i\psi_j\vert U_{ij}\vert^2
\rightarrow \nonumber
\eq
\bq
\sum_i\phi_i\psi_{P(i)}  ~+  (1/2)\sum_{i,j}\vert
H_{ij}\vert^2(\phi_i-\phi_j)(\psi_{P(i)}-\psi_{P(j)}) + ...
\label{(3.4)}
\eq
Then DH theorem implies that all the higher order terms can be ignored.
 The result of the Gaussian integration over $H$  is
\bq
\int  [dU] \exp{\left\{\sum_{i,j}\phi_i\psi_j \vert U_{ij}\vert^2\right\}}
  \nonumber
\eq
\bq
\propto ~\sum_P e^{\sum_i\phi_i\psi_{P(i)}}
\int _{-\infty}^{+\infty}\prod_{i,j} dH_{ij} e^{(1/2)\sum_{i,j}\vert
H_{ij}\vert^2(\phi_i-\phi_j)
(\psi_{P(i)}-\psi_{P(j)})}
\label{(3.5)}
\eq
\bq\propto\sum_P (-)^P
\frac{e^{\sum_i\phi_i\psi_{P(i)}}} {\Delta(\phi)\Delta(\psi)} \sim {\rm
const} ~\frac{\det_{ij} e^{\phi_i\psi_j}}{\Delta(\phi)\Delta(\psi)}
\nonumber
\eq
 where the sum is over permutations $P$ and the
 factor $(-)^P$ comes from the fact that
\bq
\prod_{i<j} (\psi_{P(i)}-\psi_{P(j)}) = (-)^P \prod_{i<j} (\psi_i-\psi_j)
 = (-)^P\Delta(\psi)
\eq
This completes the derivation of the result (\ref{(3.2)}) using the DH theorem.

Unfortunately the same methods are not directly applicable to
evaluation of the integrals $I_{pp}$ with $p\ne0$.  In fact the generating
functional for all non-vanishing integrals of this type is given
by
\bq
I[A] \equiv \int [dU] \exp \sum_{i,j} A_{ij} \vert U_{ij} \vert ^2.
\label{(3.6)}
\eq
 Unfortunately the DH formula is not
directly applicable to this case. If it were applicable then
we would find that:
\bq
I[A]\propto~\sum_P {{\exp \sum_k A_{kP(k)}}\over{\prod_{i<j}
[A_{iP(i)} + A_{jP(j)} - A_{iP(j)} - A_{jP(i)}] }}(1~+~{\cal O}(\frac{1}{A})).
\label{(3.7)}
\eq
with order $1/A$ corrections vanishing.
(Note that the  classical equations of motion are  $\sum_j U_{ij}
 (A_{ji} - A_{jk}) U^{\dagger}_{jk} = 0$. Any permutation matrix
 $U = P$ is a solution to these equations.)
Since the DH theorem is not directly applicable in this situation the
${\cal O}(\frac{1}{A})$ corrections do
not vanish in general. There are several interesting exceptions in which
case they do vanish. The first case is if
$A_{ij}$ is of rank $1$ so that $A_{ij} = \phi_i \psi_j$ (we then
recognize Eq.(\ref{(3.5)}) in Eq. (\ref{(3.7)})). Another interesting
case is when  $N=2$. For the case of $SU(2)$ one can check explicitly that
Eq. (\ref{(3.7)}) is exact so that
\bq
I_{SU(2)}[A] =\propto\frac{e^{A_{11}+A_{22}} -
e^{A_{12}+A_{21}}} {A_{11}+A_{22}-A_{12}-A_{21}}.
\label{(3.8)}
\eq
(Note that for $N=2$ the denominator is independent of $P$). Other
exceptional cases involve
certain $U$-integrals arising in the study of correlators in
multi-matrix models like
\bq
\int \int d\Phi d\Psi e^{{\rm tr}\Phi\Psi}
{\rm tr}e^{\alpha\Phi+\beta\Psi}
\label{(3.8')}
\eq
The fact that the DH theorem is  not  directly applicable in the
general case (\ref{(3.6)}) is rather obvious.
The action in the integral (\ref{(3.6)}) is not defined on any
co-adjoint orbit but on the entire group manifold which is not even
symplectic. Keeping this in mind it is not too difficult to work
out the corrected formula (see \cite{AN}) though we shall not use it
in this paper. In what follows we shall make use of the approximate formula
(\ref{(3.7)}) and of the explicit expression (\ref{(3.8)}) for $SU(2)$.

For our purpose which is the evaluation of the expectation value
 of the FWL we are mostly
interested in the correlators $I_{pp}$ with $p=1$:
\bq
   I_{ij}^{lk} = < U_{ij} U^{\dagger kl} > \equiv \int [dU]
{}~e^{\sum_{i,j} \phi_i \psi_j \vert U_{ij}\vert^2}~ U_{ij} U^{\dagger
kl} .
\label{(3'.1)}
\eq
It is useful to use a somewhat more explicit representation
of the Haar measure:\footnote{ This representation of the Haar
measure suggests a possible approach to the evaluation of the integrals
(\ref{(3.6)}). The idea is to introduce Lagrange multipliers $\lambda_{ij} =
\overline{\lambda_{ji}}$ for every one of the $\delta$-functions in
(\ref{(3'.2)}). Then Gaussian integral over $unconstrained$ $U$-variables is
easily evaluated with the result
\bq
I[A]\propto\int  \frac{\prod_{ij}^N d\lambda_{ij} e^{\sum_j \lambda_{jj}} }
{\prod_{k=1}^N {\rm det}(\lambda_{ij} - A_{ik}\delta_{ij})}.
\nonumber
\eq
For $N=2$ the integral over $d^2\lambda_{12}$ gives
\bq
\int \frac{d\lambda_{11} d\lambda_{22} e^{\lambda_{11} + \lambda_{22}}}
{\lambda_{11}(A_{21}-A_{22}) + \lambda_{22}(A_{11}-A_{12}) +
 (A_{12}A_{22}-A_{11}A_{21})} \nonumber
\eq
\bq
[\log \frac{\lambda_{11}-A_{11}}{\lambda_{11}-A_{12}} +
\log  \frac{\lambda_{22}-A_{21}}{\lambda_{22}-A_{22}}]
\nonumber
\eq
Consider the first logarithmic term. The integral over
$\lambda_{22}$ is easily evaluated yielding
\bq I[A]\propto
\frac{1}{A_{11}-A_{12}} \exp \frac{A_{11}A_{21}-A_{12}A_{22}}{A_{11}-A_{12}}
\times
\nonumber
\eq
\bq
\int d\lambda_{11} \bigl\{ \exp \lambda_{11}
\frac{A_{11}+A_{22}-A_{12}-A_{21}}{A_{11}-A_{12}} \bigr\}
\log  \frac{\lambda_{11}-A_{11}}{\lambda_{11}-A_{12}}
\nonumber
\eq
This integral can be evaluated with the help of the formula
\bq
\int d\lambda e^{c\lambda} \log \frac{\lambda-\alpha}{\lambda-\beta} =
\frac{1}{c} \int d\lambda e^{c\lambda} (\frac{1}{\lambda-\alpha} -
\frac{1}{\lambda - \beta}) = \frac{1}{c} (e^{c\alpha} - e^{c\beta})
\nonumber
\eq
and is equal to (\ref{(3.7)}). The contribution of the second logarithmic
term is is the same as the first. }
\bq
   [dU] = \prod_{i,j} dU_{ij} \delta(\sum_k U_{ik}
\overline{U_{jk}} - \delta_{ij} ) .
\label{(3'.2)}
\eq
Making use of the invariance of the measure under the transformations
\bq
   U_{ij} \rightarrow \omega^i U_{ij},~~ U^{\dagger kl}
\rightarrow U^{\dagger kl} \omega^{-l}
\label{(3'.3)}
\eq
and
\bq
   U_{ij} \rightarrow U_{ij} \omega^j,~~ U^{\dagger kl}
\rightarrow \omega^{-k} U^{\dagger kl}
\label{3'.4)}
\eq
(with $\omega^N = 1$), one can concludes that $ I_{ij}^{lk} = 0$ unless
$i=l$ and $k=j$, so that
\bq
< U_{ij} U^{\dagger kl}> \sim \delta_i^l \delta_j^k .
\label{(3'.5)}
\eq
We now introduce the quantities
\bq
    C_{ij} \equiv \frac{ < \vert U_{ij} \vert^2>}{<1>} ,~~~ i,j =
1...N ,
\label{(3'.6)}
\eq
which are the only non-vanishing quadratic correlators.  These clearly
satisfy the  relations
\bq
    C_{ij} \ge 0;~~\sum_j C_{ij} = 1 = \sum_j C_{ji}~{\rm for} \; \forall i;
{} ~~ C_{ij} [\phi, \psi] = C_{ji} [\psi, \phi],
\label{(3'.7)}
\eq
It thus follows that the  $C_{ij}$ can be interpreted as conditional
 probabilities.
These quantities will be used in the next section to evaluated the average
of the FWL.

\section{ Loop averages as partition functions of statistical systems}

   Originally the average $\ll  W(\Gamma_1 ... \Gamma_k) \gg$ for
$\Gamma_1 + ... + \Gamma_k = 0$ can be represented as a double-loop
diagram (``fat graph''). Because of the property (\ref{(3'.5)}) all
double lines can be exchanged for single lines (Fig.3) since
\bq
   \frac{ <U_{ij} U^{\dagger}_{kl} >}{<1>} = \delta_{il} \delta_{jk}
C_{ij} ,
\label{(4.1)}
\eq
Thus at every site there is only one independent value $i$ of some effective
``spin''.
These single lines are oriented because, in general, $C_{ij} \neq C_{ji}$.
 Let us
denote the graph consisting of these single lines by
$\tilde{\Gamma}$. For FWL $\tilde\Gamma = S$.  Then
\bq
   \ll  W(\Gamma_1 ... \Gamma_k) \gg ~=
\nonumber
\eq
\bq
{{\int d\phi \exp[-V(\phi)] <1>
 \sum_{i(x)}  \bigl\{
\prod_{<x,y>\in \tilde{\Gamma}} C_{i(x) j(y)} [\phi(x), \phi(y)]  \bigr\}}\over
{\int d\phi \exp[-V(\phi)] <1>} }
\label{(4.2)}
\eq
where the sum is over distributions $i(x)$ of integers $i$ at sites $x$.
The quantities on the r.h.s. are already very  similar to the
partition function of some statistical system with only nearest
neighbor interactions.  In the case of the FWL operators this system is
essentially 2-dimensional (for adjoint loop it would be
1-dimensional). It is unfortunately not a very simple theory, since
 the statistical
weights at any link depend not only on the values of the ``spin'' at
two adjacent sites, but also on the orientation of the link ($C_{ij}
\neq C_{ji}$) and on the field $\phi(x)$. We can thus think
of the statistical weights $C_{ij}$ as  random
variables with some non-trivial distribution which is encoded in terms
of some
``hidden variables'' $\phi(x)$ with gaussian distribution when  $V$
is quadratic).
Because of the presence of these random variables  the system in
question looks more like a ``spin glass'' than an ordinary spin system.
\footnote{ Because of the presence of the factor $<1>$
in (\ref{(4.1)}) which is  $\phi$ dependent, the shape of the
random distribution of the $C_{ij}$ depends on the actual choice of contour
$\Gamma$, i.e. on the form of the 2-dimensional sample $S = \tilde\Gamma$.
This is a kind of a boundary effect for the statistical system
and although it is is not very important for the derivation of results
such as the area law it is
important for the study of {\it correlators} of (filled) Wilson loops.

Note also that the r.h.s. of Eq.(\ref{(4.2)}) is a ratio  of
  averages over $\phi$-fields. Formally this is different from the
situation in conventional spin
glass systems where any back reaction of the system on the shape of the
random ``noise'' distribution is neglected and thus the average of the ratio
rather than the ratio of the average is the quantity of interest.
(In practice the difference is not so drastic, since the distribution
of the logarithms of Boltzmann weights $\log C_{ij}$ in (\ref{(4.2)})
is far from being Gaussian.)}

   A drastic simplification arises in the ``mean- (or master-) field''
approximation for the integral over $\phi(x)$ in which case  $\phi(x)$ is
frozen and does not depend on $x$, i.e. $\phi(x) = \Phi$. The $C_{ij}$ are thus
also fixed and, moreover, they are symmetric i.e. $C_{ij} = C_{ij}
[\Phi, \Phi] = C_{ji} [\Phi, \Phi] = C_{ji}$. $W(\tilde\Gamma)$ is
thus the partition function of a magnetic-type statistical system:
\bq
   \ll  W (\tilde{\Gamma}) \gg ~\sim \sum_{i(x)}~
\bigl\{\prod_{<x,y>\in \tilde{\Gamma}} C_{i(x) j(y)} \bigr\}
     \label{(4.3)}
\eq
The fascinating thing about this relationship is that the area law for
the FWL is just equivalent to the statement that
the free energy of the statistical system is proportional to its area.
This will be true provided that the free energy of this 2-dimensional
system is nonzero. This seems to be a rather obvious property for a system
with local (nearest neighbor) interactions.
There is however one delicate point which prevents
us from  using this result immediately. The problem is that
the ``Boltzman factors'' have the rather peculiar normalization (\ref{(3'.7)}).
We  now discuss the implications of this fact using the simplest
example of the Ising model which  arises for  $N=2$.

\section {The case of $N=2$: The Ising model}

When $N=2$ the diagonalized matrices $\phi$ and $\psi$ can be written as
\bq
\phi = \left( \begin{array}{cc} \hat\phi & 0 \\
 0 & -\hat\phi \end{array} \right), ~~~
\psi = \left( \begin{array}{cc} \hat\psi & 0 \\
 0 & -\hat\psi \end{array} \right)
\label{matrix}
\eq
It is now easy to express the $C_{ij}$ explicitly in terms of the
variable $\gamma=2\hat\phi\hat\psi$
One finds that
\footnote{ These formulas can be easily deduced either by an explicit
 evaluation of
the integrals over $SU(2) = S^3$ or by an application of the DH result
(\ref{(3.8)}). The simplest way to proceed with an
explicit derivation is to use the representation (\ref{(3'.2)}) for
the measure $[dU]$.
One writes $U = a_0 + i\vec{a}\cdot\vec{\sigma}$. The $\delta$-functions
in (\ref{(3'.2)})
then imply that $a_0$ and  $\vec{a}$ are real (after a  $U(1)$ factor is
extracted from the measure),
 and the measure is simply $\int d^4a \delta(a_0^2+\vec{a}^2 -1)$.
In Eq.(\ref{(5.1)}) $x \equiv a_1^2+a_2^2 = \vert U_{12}\vert^2 $ and
$1-x = a_0^2+a_3^2 = \vert U_{11} \vert^2 = \vert U_{22} \vert^2$.}
\bq
C_{11} = C_{22} = \frac{\int^1_0 (1-x)e^{-2\gamma x}dx} {\int^1_0
e^{-2\gamma x}dx} =
\frac{1-1/{2\gamma}+e^{-2\gamma}/2\gamma}{1-e^{-2\gamma}},
\nonumber
\eq
\bq
C_{12} = C_{21} = \frac{\int^1_0 x e^{-2\gamma x}dx}{\int^1_0
e^{-2\gamma x}dx} = \frac{1/2\gamma
-e^{-2\gamma}(1+1/2\gamma)}{1-e^{-2\gamma}}
\label{(5.1)}
\eq

   The case  $N$$=$$2$ is mostly distinguished by the fact, that the two
diagonal element $C_{ii}$ coincide i.e. $C_{11} = C_{22}$ (and, as a result,
$C_{12} = C_{21}$) for any pair ($\phi,\psi$). This  allows one to
represent $C_{ij}$ as
\bq
C_{11} = C_{22} = \frac{e^{J/T}}{e^{J/T}+e^{-J/T}}
\nonumber
\eq
\bq
C_{12} = C_{21} = \frac{e^{-J/T}}{e^{J/T}+e^{-J/T}}
\label{(5.2)}
\eq
so that $\ll {\cal W}(S)\gg$ is proportional to the partition function
$Z_{\rm Ising}$ of a 2-dimensional Ising model defined on the surface
$S$.  In the mean--field
approximation when $\phi(x)$ is constant,  $J$ is the same for all sites and
 this is just a conventional Ising model in a finite area $A(S)$.
More generally there is some random distribution of the
values of $J$ . This distribution will not, in general, be Gaussian
unless $T$ and $1/ \gamma$ are small. The region of small $T$ will
be of great interest to us later in this paper.
In what follows we restrict ourselves to the mean-field approximation (and,
without loss of generality we set $J=1$).

The average of the FWL can now be evaluated with the help of Eq. (\ref{(4.2)})
\bq
   \ll {\cal W}(S)\gg = \frac{Z_{\rm Ising}}{(2\cosh{(1/T)})^{2A(S)}}
\label{(5.3)}
\eq
The two dimensional Ising model is exactly soluble. We can thus
substitute the well known formula for the free energy of the Ising
model
(see, for
example, \S $141$ of \cite{LL}) to find the string tension
\bq
\alpha'[T] = \frac{-\log <{\cal W}>}{A(S)} =  \nonumber
\eq
\bq
-\frac{1}{8\pi^2} \int^{2\pi}_0\int^{2\pi}_0 d\omega_1 d\omega_2 \log
\frac{(1+x^2)^2 - 2x(1-x^2)(\cos{\omega_1}+\cos{\omega_2})}{4}
\label{(5.4)}
\eq
 where
\bq
x = \tanh{(1/T)} \nonumber
\eq

The three most interesting temperatures in the Ising model are $T=
\infty,~ T = T_{\rm crit} = 2/\log(\sqrt{2}+1)$ and $T=0$.
 $\alpha'$ is  continuously  decreasing with
decreasing  temperature $T$. At $T=0$ $\alpha'$$=$$0$
 and   at $ T = T_{\rm crit}$
$d^2\alpha'/ dT^2$ has a (logarithmic)
singularity. $T=\infty$ corresponds to
the value $\hat\Phi = \sqrt{\gamma/2} = 0$ of the mean field , while
$T=0$ corresponds to $\hat\Phi = \infty$. In terms of the original  {KMM~}
$T=\infty$ is associated with $m=\infty$ (the strong coupling limit),
while $T=T_{\rm crit}$, where long-range correlations arise in the
Ising system, should be associated with $m = m_{\rm crit}$ (the ``weak
coupling limit", where continuum--like behaviour is supposed to occur).
The problem is that at $T_{\rm crit}$  there is no reason to believe in
the validity of the mean field approximation. Our main hope is that
 as $N$ increases, the analogue of $T_{\rm crit} [N]$ moves
toward $T$$=$$0$ so that the critical behaviour of the
relevant statistical system appears in the limit $T \rightarrow 0 $,
$\phi
\rightarrow \infty$  and  $m \rightarrow m_{\rm crit}$ at which
point the mean field approximation may be trustworthy.
It is most difficult  of course to prove the existence of
a {\it second} order phase transition for all $N$.  With this guess in
mind, let us pay more attention to the point $T=0$ in the example of the
Ising model.

   The crucial feature of the point $T$$=$$0$ for our considerations is that
$\alpha' \rightarrow 0$ as $T \rightarrow 0$. \footnote{ Note,
that this result depends crucially on the normalization of the FWL operator.
If there were an extra factor of $N^{-Area} = 2^{-Area}$ in (\ref{(2.7)}),
we would get $\alpha' \rightarrow log2$ as $T \rightarrow 0$.
This would exclude any dynamics from the derivation of area law which would
be valid in the continuum limit without any need for scaling arguments. This
is already an argument in favor of our normalization choice. We shall
return to the normalization problem in  Section 6.} It thus looks
as if the string tension is vanishing in the continuum limit.
We should be more careful, however,  and consider what happens when
$T$ is small, but still non--zero. The answer for $SU(2)$ is easily extracted
from (\ref{(5.4)}) from which one finds that
\bq
\alpha' =  \frac{1}{2\hat\phi^2} +
 o(1/{\hat\phi^4}, e^{-4\hat\phi^2})
\label{(5.5)}
\eq

Approximation that $\phi$ is large near the continuum limit
is self-consistent. The integral over $\phi$ is essentially
\bq
\int d\hat\phi e^{-(m^2-m_{\rm crit}^2)\sum_x \hat\phi^2} e^{-\alpha'A(S)}
\label{(5.5.1)}
\eq
with $\alpha'$ given by (\ref{(5.5)}).
Thus when  $m^2 \rightarrow m_{\rm crit}^2$
the integral is dominated by $\phi^2 \sim 1/\sqrt{m^2-m_{\rm crit}^2}$
 which is large in this limit.

We suggest that this simple example of the Ising model where
everything can be calculated explicitly can serve as a
prototype for the  situation of $N=\infty$. We speculate that  the main
features of
this example survive in the general situation. There are, however,
several crucial differences:

{\bf (a)} The phase transition occurs at
a lower temperature $T_{\rm crit}$
which approaches $T=0$ for large $N$.

{\bf (b)} The single eigenvalue $\hat\phi$ gets
replaced by a distribution of eigenvalues with a density $\rho(\phi)$.}

{\bf (c)} The  $\exp (-\hat\phi^2)$--type corrections can become relevant for
 smooth functions $\rho(\phi)$.

 If this suggestion is correct then we would expect
Eq. (\ref{(5.5)}) to be replaced by
\bq
\alpha' \propto \int\int \frac{\rho(\phi)\rho(\phi')d\phi d\phi'}
{(\phi-\phi')^2}
\label{(5.6)}
\eq
(note that $2\hat\phi$ in (\ref{(5.5)}) is just the difference
$\phi_1-\phi_2$ of two eigenvalues of (\ref{matrix}).
It  thus follows that $\alpha'$ is not necessarily
vanishing in the continuum limit. Critical behaviour would arise at
large but finite values of the master field $\Phi$.

 \section{ Puzzles of the continuum limit of  KMM}

Before we turn to the general analysis of the $SU(N)$ case, let us
examine  in more detail the two limiting cases of strong and weak coupling.

   The strong coupling limit of the model corresponds to $m
\rightarrow \infty$. The $\exp -m^2 \phi^2$ factor in the
integrand then damps all  fluctuations of the $\phi$ fields, which are thus
restricted to vanish. It follows that
 the action in the integral (\ref{(3'.1)}) over $U$ vanishes
and what remains in the evaluation of expectation values
 are just integrals with the conventional Haar measure, of the form
\bq
\frac{<U_{ij}U^{\dagger}_{kl}>}{<1>}\vert_{\phi=0} =
\frac{\int[dU] U_{ij}U^{\dagger}_{kl}}{\int [dU]} =
\frac{1}{N} \delta_{il}\delta_{jk}
\label{(6.1)}
\eq
so that
\bq
C_{ij}[\phi=\psi=0] = \frac{1}{N}
\eq
is equal for all $i$ and $j$.
(One can similarly derive expressions such as
\bq
\frac{<U_{ij}U_{mn}U^{\dagger}_{kl}U^{\dagger}_{pq}>}{<1>}\vert_{\phi=0} =
\frac{1}{N^2-1}(\delta_{il}\delta_{jk}\delta_{mq}\delta_{np} +
\delta_{iq}\delta_{jp}\delta_{ml}\delta_{nk}) - \nonumber
\eq
\bq
\frac{1}{N(N^2-1)}(\delta_{jk}\delta_{lm}\delta_{np}\delta_{qi} +
\delta_{jp}\delta_{qm}\delta_{nk}\delta_{li}).
\label{(6.2)}
\eq
 In this strong coupling approximation any average of a gauge invariant
product of Wilson loops $(2.4)$ is simply equal to
\bq
\ll  W(\Gamma_1...\Gamma_k) \gg\vert_{strong ~coupling} = \frac{1}{N}
\frac {N^{\#  vertices }}{N^{\# links}}
= \frac{1}{N^{\#  loops}} = \frac{1}{N^{k-1}}
\label{(6.3)}
\eq
provided $\Gamma_1 + ... + \Gamma_k = 0$.
For the FWL operator $k-1 = \#$(of plaquettes in S)  $= A(S)$, and we
obtain:
\bq
\ll  W(\Gamma_1...\Gamma_k) \gg\vert_{\rm strong ~coupling} =
\frac{1}{N^{Area}}
\label{(6.4)}
\eq
The area law is thus trivially true in the strong coupling limit. Of
course, the really interesting question is whether it is also true in
the ``weak coupling" (continuum) limit.

   Naively the ``weak coupling limit" corresponds to $m
\rightarrow 0$, where integrals over $\phi$ are dominated by large
values of the $\phi$-fields (i.e. all the eigenvalues $\phi_i$ are large).
In fact integrals over $\phi$ diverge already for a finite value of
$m = m_{\rm crit}$ (quasi--classically $m_{\rm crit}^2 = D$), and the
relevant limit is $m \rightarrow m_{\rm crit} + 0$. The fact that
integrals over $\phi$ diverge still allows  interesting quantities
such as  $<\vert U_{ij}\vert^2 > \over <1>$, which are ratios of divergent
integrals, to be well defined. In fact, when all $\phi_i$'s and
$\psi_i$'s are large (and positive) the integral
\bq
\int[dU] \exp{\sum_{i,j}\phi_i\psi_j\vert U_{ij}\vert^2}
\label{(6.5)}
\eq
is clearly dominated by diagonal matrices
\bq
U_{ij} = e^{i\theta_i}\delta_{ij}, ~~ \vert U_{ij}\vert^2 = 1
\label{(6.6)}
\eq
and thus
\bq
\frac{<U_{ij}U^{\dagger}_{kl}>}{<1>}~\stackrel{\phi,\psi \gg1}{\rightarrow}
\delta_{il}\delta_{jk}\delta_{ij},~~~~
C_{ij} ~\stackrel{\phi,\psi \gg1}{\rightarrow} \delta_{ij}
\label{(6.7)}
\eq
In this approximation it is clear that all the spins in
the  statistical system take  on the same value at each site.
(This is clearly the ordered phase which occurs in the spin system at
zero temperature.) There is just an overall degeneracy factor
of $N$, which cancels the only $N$ in the normalization in
(\ref{(2.7)}):
\bq
\ll  W(\Gamma_1...\Gamma_k) \gg\vert_{m=m_{crit}} =
1\times\delta(\Gamma_1+...+\Gamma_k)
\label{(6.8)}
\eq
In particular in this approximation the FWL is given by
\bq
\ll  {\cal W}(S) \gg\vert_{m=m_{crit}} = 1
\label{(6.9)}
\eq
and, as explained at the end of the previous section, the vicinity of
the point $m = m_{\rm crit}$ should be investigated in order to
explain whether and why $\alpha'$ may be non-vanishing in the continuum limit.

 This will be our purpose in the next section. In this section we
discuss what happens in the ``naive continuum limit". As we already
mentioned in Section 2, evaluation of ${\cal W}(S)$ in the naive
continuum limit would give a factor  $N^{k-1}$ instead of $1$
(there is $k-1$ instead of $k$ in the exponent is because of
normalization factor $1/N$ in (\ref{(2.7)})). This estimate
was however based on the implicit suggestion that
\bq
\frac{<U_{ij}U^{\dagger}_{kl}>}{<1>}\vert_{ncl} = \delta_{ij}\delta_{kl}
\label{(6.10)}
\eq
which is obviously inconsistent with the basic property (\ref{(4.1)})
of the correlators of $U$ (which can of course be traced back to the
$Z(N)$ gauge symmetry). This argument is already sufficient to demonstrate the
failure of naive continuum limit.  We shall now present a somewhat more
constructive argument against the naive continuum limit.

The discrepancy between the naive continuum limit
(which we claim is {\it incorrect} for the KMM) and the limit of $m
\rightarrow m_{\rm crit}$ for the eigenvalue model (\ref{(1.7)})
(which we claim is the {\it correct} prescription for the KMM) appears already
in a very simple situation, which can be examined {\it exactly}.
Indeed consider a lattice which is just a closed 1-dimensional chain
of $L$ links. All the techniques and results which we
derived for a generic D-dimensional lattices is
applicable to this simplified model. The set of (gauge-invariant)
observables is very small. In fact only adjoint Wilson loops and ``Baryon''
loops
 along the closed
chain and their ``powers" (the same contour passed several times) are nonzero.
Such adjoint Wilson loops are given by
\bq
{\cal O}_k = \frac{1}{N} \vert W(k\Gamma) \vert^2 .
\label{(6'.1)}
\eq
The predictions of our above analysis are that in the strong coupling limit
\bq
\ll {\cal O}_k\gg\vert_{m=\infty} = \frac{1}{N} \frac{N^{\#~{\rm of~
vertices}}}{N^{\#~{\rm of~links}}} =
\frac{1}{N} \frac{N^{kL}}{N^{kL}} = \frac{1}{N}
\label{(6'.2)}
\eq
while in the weak coupling limit (near $m_{\rm crit}$)
\bq
\ll {\cal O}_k\gg\vert_{m=m_{\rm crit}} =
\frac{1}{N}\cdot N =1~~~~~{\rm in~the~limit~}~m\rightarrow m_{\rm crit}
\label{(6'.3)}
\eq
On the other hand the ``naive continuum limit" prescription (\ref{(6.10)})
would give
\bq
 \ll {\cal O}_k\gg\vert_{ncl} = \frac{1}{N} ({\rm tr}I)^2 = N,
\label{(6'.4)}
\eq
which differs from (\ref{(6'.3)}).

We shall now  solve the original  KMM (\ref{(1.1)}) exactly for this
particular lattice and prove that while (\ref{(6'.2)}) and
(\ref{(6'.3)}) are correct predictions the ``prediction'' (\ref{(6'.4)}) is
{\it
incorrect}. We shall also find a  reason for the failure of
(\ref{(6'.4)}) which can serve as a general explanation of why the
naive continuum limit is wrong in the KMM in higher dimensions.
Recall that the
argument in support of (\ref{(6'.4)}) was that the action in
(\ref{(1.4)}) diverges as $m \rightarrow m_{\rm crit}$ and it
 has a single maximum  as a function of $U$ which occurs when
$U=I$. This is absolutely true but what is unusual is that
in the KMM is that the
peak of the action at $U=I$ is very moderate. In fact the action grows
only
logarithmically as $U$ approaches $I$.  This makes it necessary to
take the measure of integration $[dU]$ into consideration. The
 vicinity of $U=I$ where action is big has, of course, a small
volume and it  appears that this is
enough to compensate completely for the peak of
the action. In the particular example of the
one dimensional chain  we shall see
that when $m = m_{\rm crit}$ the action and the measure  cancel each
other
 completely over the entire
range of integration over $U$.  Once
observed, this phenomenon is clearly a very general feature of the
KMM. The integrand of the partition function  is very smooth as the
continuum
limit is approached so that
even quasi--classically it has is no dominant peak.
The continuum limit is thus far from naive.

Let us now turn to our calculation for the one dimensional chain.
The quantity to
compute is
\bq
\ll {\cal O}_k\gg ~ \equiv \nonumber
\eq
\bq
\frac{1}{Z}\int \prod_{x = 1}^L d\Phi_{x}[dU_{x,x+1}]
 e^{-m^2{\rm tr}\Phi_{x}^2 ~ +~{\rm tr}~\Phi_{x} U_{x,x +1}\Phi_{x+1}
U_{x,x +1}^{\dagger}}
\vert {\rm tr} [\prod_{x=1}^L U_{x,x +1}]^k \vert^2
\label{(6'.5)}
\eq
with
\bq
Z={ \int \prod_{x = 1}^L d\Phi_{x} [dU_{x,x+1}]
e^{-m^2{\rm tr}\Phi_{x}^2 ~+~
{\rm tr}~\Phi_{x} U_{x,x+1}\Phi_{x+1} U_{x,x +1}^{\dagger}} }
\nonumber
\eq
($x$$=$$L+1$ is identified with $x$$=$$1$). This one dimensional case
is particularly simple since it is possible to perform a gauge transformation
(\ref{(2.2)})  which eliminates (i.e. makes equal to
$I$) all the matrices $U_{x,x+1}$ except for a single one,
say $U\equiv U_{L,1}$. In other words
\bq
Z=\int [dU] \prod_{x=1}^L \int d\Phi_{x} e^{-m^2{\rm tr}\Phi_{x}^2}
\prod_{x =1}^{L-1} e^{{\rm tr}\Phi_{x}\Phi_{x+1}}
e^{{\rm tr}\Phi_LU\Phi_1U^{\dagger}}
\label{(6'.6)}
\eq
and $\ll {\cal O}_k\gg $ is simply the normalized average
of $\vert {\rm tr}U^k \vert^2$ with this generating functional.

The next step is to integrate first over
$\Phi_2,...,\Phi_{L-1}$ and then over $\Phi_1$. All these integrals are
 Gaussian and are thus easy to perform.
The result can be expressed in terms of the determinant $\mu_l$ of
the $l\times l$ matrix
\bq
{\cal M}_l \equiv
 \left( \begin{array}{cccccc}
 2m^2 & -1 & 0 &  \; &  0 & 0 \\
                    -1 & 2m^2 & -1 &~\cdots~    & 0   & 0  \\
                     0 &   -1 & 2m^2 &    \;        & 0 & 0  \\
                      \,&\,&\,&\cdots & \,& \,\\
                      0 & 0 & 0 & \, & 2m^2 & -1 \\
                     0 &    0  &  0 &  ~\cdots~   &-1 & 2m^2
\end{array} \right)
\eq
The determinant $\mu_l \equiv {\rm det}
{\cal M}_l$
 satisfies the recursion relation   $\mu_{l+1} = 2m^2\mu_l -
\mu_{l-1}$ with ``initial conditions" $\mu_0=1$,
$\mu_1=2m^2$ (in this respect they are somewhat similar to Fibonacchi
numbers $f_{l+1}$$=$$
f_l + f_{l-1}$, $f_0$$=$$f_1$$=$$1$).
Since the matrix ${\cal M}_l$ is simply a Laplacian the $\mu_l$  can
also be represented as
\bq
\mu_l = 2^l \prod_{k=1}^l (m^2 - \cos \frac{\pi k}{l+1})
= \nonumber
\eq
\bq
\frac{(1-q)(1-\nu q)\cdots (1-\nu^{2l+1}q)}{q^l(1-q^2)}
=\frac{1-q^{2l+2}}{q^l(1-q^2)}
\eq
where
\bq
q = q_{\pm} \equiv m^2 \pm \sqrt{m^4-1}
\label{(q)}
\eq
and $\nu = \nu_l \equiv e^{\frac{2\pi i}{2l+2}}$.

We now do the integral over $\Phi_x$ in
(\ref{(6'.6)}):
\bq
\int d\Phi_1 d\Phi_L e^{-m^2{\rm tr}(\Phi_1^2 + \Phi_L^2)}
e^{{\rm tr}\Phi_L U \Phi_1 U^{\dagger}}~\prod_{x=2}^{L-1} \int d\Phi_x
 e^{-m^2{\rm tr} \Phi_x^2} e^{{\rm tr}\Phi_1\Phi_2 +
{\rm tr}\Phi_x\Phi_{x+1}}
\label{(6".2)}
\eq
$\Phi_1$ and $\Phi_L$ can be considered as ``source terms'' in the Gaussian
integral over  $\Phi_2,...,\Phi_{L-1}$ with the action
\bq
\frac{1}{2}\sum_{x,y=2}^{L-1}
 {\rm tr}\Phi_x ({\cal M}_{L-2})_{xy} \Phi_y =
\label{(6".3)}
\eq
\bq
\mu_{L-2}^{-N^2/2} \exp{\frac{1}{2}
[\Phi_1^2({\cal M}_{L-2}^{-1})_{22} +
2\Phi_1\Phi_L({\cal M}_{L-2}^{-1})_{2,L-1} +
\Phi_L^2({\cal M}_{L-2}^{-1})_{L-1,L-1}]}
\nonumber
\eq
The relevant matrix elements are
\bq
({\cal M}_{L-2}^{-1})_{22} =
({\cal M}_{L-2}^{-1})_{L-1,L-1}=\frac{\mu_{L-3}}{\mu_{L-2}}
\nonumber \\
  {\rm and~~~~}({\cal M}_{L-2}^{-1})_{2,L-1}=\frac{1}{\mu_{L-2}}
\eq
 Since ~~$m^2 - \mu_{L-3}/2\mu_{L-2}
=\mu_{L-1}/2\mu_{L-2}$ ~~we get for (\ref{(6".2)}):
\bq
 \mu_{L-2}^{-N^2/2}
\int d\Phi_1 d\Phi_L \exp[-\frac{\mu_{L-1}}{2\mu_{L-2}}{\rm tr}
(\Phi_1^2 + \Phi_L^2) +  \nonumber
\eq
\bq
\frac{1}{\mu_{L-2}}
{\rm tr}\Phi_1\Phi_L + {\rm tr}\Phi_L U \Phi_1 U^{\dagger}].
\label{(6".4)}
\eq
The remaining integration over $\Phi_1$ is done  by rescaling  $\Phi_L
\rightarrow \Phi \equiv \Phi_L/\sqrt{\mu_{L-1}}$
The result is:
\bq
Z~ \sim  {\cal N}_L
\int [dU] \int d\Phi ~e^{-M_L^2{\rm tr}
\Phi^2~ +~{\rm tr}\Phi U \Phi U^{\dagger}}
\label{(6'.10)}
\eq
with ${\cal N}_L = 1$ and
``renormalized" (generically lattice and contour-dependent)
mass parameter
\bq
M_L^2 = \frac{\mu_{L-1}^2 - \mu_{l-2}^2 - 1}{2\mu_{L-2}} =
\frac{\mu_{2L-2}-1}{2\mu_{L-2}}
=\frac{1+q^{2L}}{2q^L}~~.
\label{(6".5)}
\eq
In terms of the new parameter
$q_L \equiv M_L^2 - \sqrt{M_L^4-1}$ (i.e. $M_L^2 = (1+q_L^2)/2q_L$),
Eq.(\ref{(6".5)}) turns into a simple relation:
\bq
q_L = (q_{\pm})^{\mp L}
\eq
where $q_{\pm}$ is defined in (\ref{(q)}).

The  integral for $Z$ and for $\ll {\cal O}_k\gg $ is easily
 evaluated
by performing a gauge rotation
on $U$ and $\Phi$
so as to make $U$ diagonal ($U = {\rm
diag}(e^{i\theta_1},...,e^{i\theta_N})$). It follows that
\bq
{\rm tr}U^{k} = \sum_{j=1}^N e^{ik\theta_j}
\nonumber
\eq
 while the Haar measure is
\bq
[dU] \sim \prod_{j=1}^N d\theta_j ~\Delta^2(e^{i\theta}) =
\prod_{j=1}^N d\theta_j \prod_{i<j}^N (e^{i\theta_i}-e^{i\theta_j})^2~
\sim  \nonumber \\
{}~ \prod_{j=1}^N d\theta_j \prod_{i<j}^N \sin^2\frac{\theta_i-\theta_j}{2}.
{}~~~~~~~~~~~~~~~~~~~~~
\label{(6'.12)}
\eq
For diagonal $U$ the action in (\ref{(6'.10)}) becomes
\bq
\sum_{i,j}^N
\vert\Phi_{ij}\vert^2(M_L^2\delta_{ij} - e^{i(\theta_i-\theta_j)}) =
\nonumber
\eq
\bq
(M_L^2-1)\sum_i \Phi_{ii} +
2\sum_{i<j}\vert\Phi_{ij}\vert^2(M^2-\cos(\theta_i-\theta_j))
\eq
The
integral over $\Phi$ is now easily evaluated.  Using
(\ref{(6'.12)}) one finds
\bq
Z~
\sim  \frac{1}{M_L^{N^2-N}(M_L^2-1)^{N/2}}
\int  \prod_j^N d\theta_j \prod_{i<j}^N \frac{\sin^2
\frac{\theta_i-\theta_j}{2}}
{1~-~\frac{\cos(\theta_i-\theta_j)}{M_L^2}}
=  \nonumber
\eq
\bq
= 2^{N^2/2}\int \prod_j^N d\theta_j
\prod_{i<j}^N \sin^2\frac{\theta_i-\theta_j}{2}
\prod_{i,j}^N \frac{q^{L/2}}{1~-~q^Le^{i(\theta_i-\theta_j)}}
\label{(6'.13)}
\eq
The critical value of $M$ is  clearly
$M_{\rm crit} = 1$.\footnote{Eq.(\ref{(6".5)}) implies
that $M_{\rm crit}^2 = 1$ ($q_{\rm crit} = 1$) simply
corresponds $m_{\rm crit}^2 = 1$,
which is the same for all
values of L. (This is easy
to check, since for $m^2=1$ the formula for $\mu_l$ is very simple:
$\mu_l = l+1$.) This value of $m_{\rm crit}^2$ is also  in
accordance with the simple quasi--classical estimate
$m_{\rm crit}^2 = D$  (see footnote 3
above).

Eq.(\ref{(6'.13)}) was also deduced in \cite{BK}
 and in a very recent paper \cite{CAP},
where also the result of integration over all $\theta$-variables
was derived:
\bq
(\ref{(6'.6)}) \sim (\ref{(6'.13)}) \sim \frac{q_L^{N^2/2}}{(1-q_L)
(1-q_L^2)\cdots (1-q_L^N)};~~~~~  q_L=(q_{\pm})^{\mp L}.
\nonumber
\eq
}
At this critical point
the integrand in (\ref{(6'.13)}) becomes equal to unity. The Jacobian
factor in
the measure,
$\prod_{i<j} \sin^2\frac{\theta_i-\theta_j}{2}$ completely cancels the
contribution from the action, $\prod_{i<j} (1-\cos(\theta_i-\theta_j))$.
Note that the
contribution  of the action is  in in denominator and is indeed
singular as $M \rightarrow M_{\rm crit}$ but, as stated above,
the singularity is
very moderate.  We may now evaluate the
average of $\vert {\rm tr}U^k \vert^2$ at $M = M_{\rm crit}$. Since
the measure of integration is  just $ \prod_j^N d\theta_j $ we find
\bq
\ll {\cal O}_k\gg\vert_{m=m_{\rm crit}} = \frac{1}{N} \frac {\int
\prod_j^N d\theta_j \vert \sum_{j=1}^N e^{ik\theta_j} \vert^2}{\int
\prod_j^N d\theta_j} = \frac{N}{N} = 1
\label{(6'.14)}
\eq
in accordance with (\ref{(6'.3)}) (and contrary to
(\ref{(6'.4)})).
This completes the proof. The conclusions which we deduce from this
result have been discussed previously.

 In the remainder of this section we  use the opportunity
provided by this calculation to make  another important  observation.
Note, that there is a somewhat different way to evaluate the integral
(\ref{(6'.10)}). It can be represented as the determinant of an $N^4\times
N^4$ matrix,
\bq
(\ref{(6'.10)}) ~~ \sim ~~ {\rm det}^{-1/2} \bigl\{ M^2I\otimes I -
\frac{1}{2}(U\otimes U^{\dagger} + U^{\dagger}\otimes U) \bigr\}~ \sim
\nonumber
\eq
\bq
M^{-N^2/2}\exp \bigl\{ -\frac{1}{2} {\rm tr}\log
[I\otimes I -
\frac{1}{2M^2}(U\otimes U^{\dagger} +
U^{\dagger}\otimes U)] \bigr\} =
\eq
\bq
M^{-N^2/2}\exp \frac{1}{2}
\sum_{l=1}^{\infty}
 \frac{1}{lM^{2l}}{\rm tr}(\frac{U \otimes U^{\dagger}
+ U^{\dagger}\otimes U}{2})^l \nonumber
\label{(6".11)}
\eq
If we now substitute $U$ in the diagonal form
this expression becomes
\bq
M^{-N^2/2}\exp \frac{1}{2} \sum_{l=1}^{\infty}\frac{1}{lM^{2l}}
\sum_{i,j=1}^N (\frac{e^{i(\theta_i-\theta_j)} +
e^{-i(\theta_i-\theta_j)}}{2})^l = \nonumber
\eq
\bq
M^{-N^2/2}\exp \{ \frac{N}{2} \sum_{l=1}^{\infty}\frac{1}{lM^{2l}}
{}~+~  \sum_{i<j}^N  \sum_{l=1}^{\infty}\frac{1}{l}
(\frac{\cos (\theta_i-\theta_j)}{M^2})^l \bigr\} ~=~
\eq
\bq
\frac{1}{M^{N^2-N}}\frac{1}{(M^2-1)^{N/2}}
\frac{1}{\prod_{i<j}^N(1~-~\frac{\cos (\theta_i-\theta_j)}{M^2})}\nonumber
\label{(6".12)}
\eq
i.e. we reproduce the  contribution of the action to (\ref{(6'.12)}).

The exponent at the r.h.s. of (\ref{(6".11)}) can now be
rewritten as
\bq
\frac{1}{2} \sum_{l=1}^{\infty} \frac{1}{l(M^2)^l} \sum_{k=0}^l
\frac{l!}{2^lk!(l-k)!}\vert{\rm tr}U^{l-2k}\vert^2.
\label{(6".13)}
\eq
In this expression one can easily recognize the $D=1$ version
of (\ref{(1.4)}).

We now review our conclusions from this section.
 Our main claim is that the action grows only
{\it logarithmically} in the vicinity of the critical point
$m=m_{\rm crit}$. This fact is of course related to the occurrence of the
factor $1/l$ in the sum over Wilson loops in
Eqs.(\ref{(1.4)},\ref{(6".13)}). However, the paths
 which are summed over in these formulae  are allowed to have
 steps which ``double back'' on themselves  (Fig.4).
Sums over such inverse steps introduce
complicated combinatorial factors such as the binomial coefficients in
(\ref{(6".13)}). The above calculation serves as an illustration that
these complicated corrections {\it do~not} necessarily destroy the
logarithmic  behaviour but rather renormalize the bare mass parameter.

\section{The area law in the case of arbitrary N}

So far we have examined two exactly solvable examples. We have
solved the KMM for the group $SU(2)$  and  for
1--dimensional lattices. We return now to the general case
where $N$ is arbitrary and the lattice is  $D$--dimensional for which
an  exact solution is not yet  available. We now suggest a qualitative
picture  which we expect to be valid in the generic situation.
This would include  the following ingredients.

{\bf (a)} Gauge invariant observables in the KMM are associated with
either one or two dimensional sublattices $\tilde\Gamma$ of original
lattice. The corresponding average is given by the partition function
of a $1d$ or of a $2d$ statistical system on $\tilde\Gamma$
(which is defined by the Boltzmann weights $C_{ij}[\phi,\psi]$ with
randomly distributed $\phi,\psi$). FWL operators are associated with
two dimensional lattices $\tilde\Gamma$ and thus with  two dimensional
statistical systems.

{\bf (b)} The area law for the average of the FWL is simply a reflection
of the fact that the free energy of a $2d$ statistical system is
proportional to its size (area).

{\bf (c)} In the strong coupling limit  $m \rightarrow \infty$
the string tension $\alpha' = \frac{1}{a^2}\log N$.

{\bf (d)} Long range correlations  occur in $2d$ statistical system,
if they have  a second order phase transition, at a certain
temperature $T[N;m] = T_{\rm crit}[N]$ corresponding to
a certain value of $m = \tilde m_{\rm crit}[N]$. It seems that
a necessary condition for  the KMM to have a continuum limit is
the presence of long range correlations in the $2d$ subsystem and
 thus  the phase transition at $T_{\rm crit}$ must be
second order.

{\bf (e)} The partition function of the $2d$ statistical system can
easily be evaluated in the vicinity of an a--priori different
point $T[N;m] = 0$ when $m = m_{\rm crit}[N]$.
(The partition function of the KMM diverges at this point.
This is why the subscript ``crit" seems reasonable.) When
$T$ is precisely equal to zero the string tension $\alpha' = 0$.
However, in the vicinity of the point $T=0$ it obeys some scaling law
such as $\alpha' \sim \ 1/a^2N^{\gamma}$.
It can thus be finite in the appropriate limit $a \rightarrow 0,
N \rightarrow \infty$. The region of small temperatures is easy to
analyze since fluctuations of the Boltzmann weights are suppressed i.e.
the $\phi$-fields are ``frozen''
and the  statistical system is in its ``ordered" phase.

{\bf (f)} It may be hard to find the critical  temperature $T_{\rm crit}[N]$
for the actual
 system (\ref{(4.2)}). If, however, (\ref{(4.2)}) is evaluated in
 the (unjustifiable)
 approximation of frozen $\phi$-fields then $T_{\rm crit}[N] > 0$,
   and, moreover,
the phase transition is generically first order. However, in the limit
$N \rightarrow \infty$ both $T_{\rm crit}[N] \rightarrow 0$ and
the latent heat $\Delta \rightarrow 0$. The reason for this is that  the
number of degrees of freedom and thus the fluctuations are increasing as
$N \rightarrow \infty$. This increase in the fluctuations should also occur
for finite $N < \infty$ if one goes beyond the frozen--$\phi$ approximation.
Thus in the large $N$ limit  the two interesting values of $T$ namely
 $T=T_{\rm crit}$ from d) and $T=0$ from e) seem to approach each
other. Thus the continuum limit of the KMM (if it exists at all)
can be examined in terms of the phase transitions of $2d$ statistical
  systems.  We thus expect the continuum limit  to occur at
  small ``temperatures''.

{\bf (g)} Our description of the continuum limit is somewhat more intricate
than the usual description. This is a peculiarity of the KMM where the
``naive continuum limit"  which is used in the Wilson theory
($\int [dU]\exp  (\frac{N}{g^2}\sum W(\Box)+ c.c.)$) and in the ``adjoint''
 theory ($\int [dU] \exp (\frac{1}{m^2}\sum \vert W(\Box) \vert^2)$)
 ($without$ a
sum over non-elementary contours!)) $does ~not ~exist$. Although the
  most
natural assumption after the failure of the ``naive continuum limit''
would be that there is no continuum limit at all, the above scenario
implies that it still can exist  but that the approach to this limit
is more sophisticated and interesting in that it is described in terms
of $2d$ systems!

The remainder of this section is devoted to various comments on this
 scenario.
Most of them concern notes d)-f) above. The note g) was already
 discussed in
some detail (though not exhaustively) in Section 6. Notes a)-c)
 were discussed
in Sections 2 and 4 and illustrated by the example of the Ising model
 in Section 5.
Our goal now is to try to generalize some results of that section to
 $N>2$.

We  begin by deriving the literal analogue of Eq.(\ref{(5.5)})
for the
average of the FWL for the case of generic $N$. We shall subsequently
discuss  the
validity of the result. For this derivation we simply need  to find
the first
order correction to the free energy of our statistical system at
low temperature in
the master-field approximation for $\phi$ assuming that the system
is in its ordered phase. Since at $T=0$  $C_{ij} = \delta_{ij}$ we introduce a
new variable $\xi_{ij}$ where $C_{ij} = \delta_{ij} + \xi_{ij}$. In the ordered
phase the ``spins'' at all the sites of the sublattice $\tilde
\Gamma = S$ have the same value $i$. Taking into account the elementary
fluctuations (i.e. the flip
of a ``spin''  $i \rightarrow j$ at one point of the lattice) as well as the
deviation of $C_{ii}$ from unity, we get:
\bq
\alpha' = -\frac{\log \ll {\cal W}(S) \gg }{a^2A(S)} ~ =
\nonumber
\eq
\bq
 -\frac{1}{a^2 NA(S)}
\sum_{i=1}^N \left( \sum_{{\rm links} \in S} (\xi_{ii} +
 \sum_{j\neq i}\xi_{ij}^{2 D}) \right) +
{\rm higher~order~corrections}
\label{(7.1)}
\eq
Due to the normalization condition (\ref{(3'.7)})
$\xi_{ii} = -\sum_{j\neq i} \xi_{ij}$,
while the second term
in (\ref{(7.1)}) can be neglected since $2D > 1$. Thus
\bq
\alpha' = \frac{2}{a^2N} \sum_{i\neq j} \xi_{ij}.
\label{(7.2)}
\eq
In order to estimate $\xi_{ij}$ in the limit of large $\phi$
(which is the same as ``low temperature'' and
$m \rightarrow m_{\rm crit}$), one can make use
of the formula (\ref{(3.7)})  (${\cal O}(\frac{1}{A})$
corrections being unessential in this limit).

The leading contribution to (\ref{(3.7)}) comes from the identity
permutation $P=I$.
The corresponding contribution to $\xi_{ij}$ is
\bq
\frac{\partial}{\partial A_{ij}} \log \frac{e^{\sum_kA_{kk}}}
{\prod_{k<l}(A_{kk}+A_{ll}-A_{kl}-A_{lk})} \left( 1~+~{\cal O}
\left(\frac{1}{A} \right) \right)\vert_{A_{ij}=\phi_i\phi_j}
\label{(7'.1)}
\eq
Since we are only interested in the case $i\neq j$, this gives
\bq
\frac{1}{A_{ii}+A_{jj}-A_{ij}-A_{ji}}
 \left( 1~+~{\cal O}
\left(\frac{1}{A} \right) \right)\vert_{A_{ij}=\phi_i\phi_j} =
\nonumber
\eq
\bq
\frac{1}{(\phi_i-\phi_j)^2 }\left( 1~+~{\cal O}(\frac{1}{\phi^2}) \right)
\label{(7'.2)}
\eq
In the approximation where all the $\phi$'s are large and {\it different},
the ${\cal O}\left(1/\phi \right)$--corrections can be neglected.
The same is true about the contributions of other permutations
$P\neq I$, which are
suppressed exponentially: the contribution of $P_{kl}$, which permutes
$k$ and $l$, is ${\cal O}\left( e^{-(\phi_k - \phi_l)^2} \right)$.
Therefore the literal analogue of (\ref{(5.5)}) for generic $N$ is
\bq
\alpha' = \frac{2}{a^2N} \sum_{i \neq j}\frac{1}{(\phi_i-\phi_j)^2}
\left( 1~+~{\cal O}(\frac{1}{\phi^2};~e^{-(\phi_k-\phi_l)^2}) \right)
\label{(7'.3)}
\eq

However as $N \rightarrow \infty$ the most interesting phase,
associated with the continuum limit of the KMM, seems to occur when
eigenvalues $\phi_i$, though large ($\phi_i \gg 1$), are not all very
 different i.e. $\frac{( \phi_i-\phi_j)^2}{\phi_i^2} \ll 1$. In other
words
the eigenvalues should be smoothly distributed with some density
 $\rho(\phi)$ ($\int \rho(\phi)d\phi = N$),
so that $\sum_i F(\phi_i) \sim  \int \rho(\phi)d\phi F(\phi)$.
If expressed
in terms of $\rho(\phi)$, Eq.(\ref{(7'.3)}) looks like
\bq
\alpha' = \frac{2}{a^2N}\int \frac{\rho(\phi)\rho(\phi')d\phi
d\phi'}{(\phi -\phi')^2} + {\rm corrections}.
\label{(7'.4)}
\eq

For smooth distributions $\rho(\phi)$ the corrections,
neglected in (\ref{(7'.1)}), can become essential.
First of all there can be corrections which contain negative
powers of the {\it differences} $(\phi_i-\phi_j)$ which are enhanced in
integrals like (\ref{(7'.4)}). Indeed the
${\cal O}(1/A)$-corrections
in (\ref{(3.7)}) contain terms with negative
powers $(A_{iP(i)}+A_{jP(j)}-A_{iP(j)}-A_{jP(i)})^{-k}
\rightarrow (\phi_i-\phi_j)^{-2k}$ with $k=1\cdots {\rm rank}(SU(N))$.
If it were
not the $k=1$ term (as supposed in (\ref{(7'.1)})) but the $k=N-1$ term
which dominated then we would get an extra factor of $(N-1)$ on the r.h.s. of
(\ref{(7'.3)}). It may however happen that the contributions of all $k$ are
equally important in which case the
exact counterpart of (\ref{(3.7)}) should be
used (see \cite{AN}).
Secondly when $(\phi_k-\phi_l)^2$ is not usually large the exponential
corrections, (arising from non-trivial permutations $P$), are also
important. Expressed in terms of the
density $\rho(\phi)$ they have the form
\bq
\int e^{(\phi-\phi')^2}G(\phi,\phi') \rho(\phi)\rho(\phi')d\phi d\phi'
\eq
where $G$ is some rational function of $\phi$ and $\phi'$.

We shall not go into more detail about the evaluation of these corrections
 because the very approximation (the low temperature
expansion for the ordered
phase) may be unreliable in the situation of interest.
As  we have stated previously,
in order for a non--trivial continuum limit to occur there
are several different phenomena which must occur simultaneously.

{\bf (a)} The auxiliary 2-dimensional statistical system
must have long range
correlations. This is most probably realized at the critical
point $T_{\rm crit}$
which should be (nearly) second order.

{\bf (b)} The master-field approximation for the
$\phi$-fields must be applicable. This
is a reasonable approximation when $N \rightarrow \infty$ and
it leads
to the occurrence of magnetic-type statistical systems.

{\bf (c)} The ``temperature'' $T$ which characterizes
the behaviour of the 2-dimensional
system should be close to zero.  At
$T=0$ the string tension vanishes, $\alpha'=0$. This is a normalization
condition for the ``Boltzman weights" $C_{ij}$ related to (\ref{(3'.7)}).
For the statistical system it corresponds to subtracting the ground state
energy and is independent of other details of the system.  On the other
hand the area law which occurs at small but non-vanishing temperature
depends on the the actual dynamics of the statistical system,
i.e. on whether there is a scaling limit at $T\sim0$.

These requirements are mutually consistent at large $N$, when
$T_{\rm crit}[N]$
is assumed to approach zero. This in turn implies that $T=0$ is in
fact a critical point and thus one cannot rely on
the low temperature expansion  in the vicinity
of this point (as we have done for
 the Ising model). One should go beyond this approximation and apply
some formalism
adequate for the study of phase transitions. This should be,
however, a
solvable problem, since it concerns a phase transition in a 2-dimensions
(though the KMM itself is formulated in $D$ dimensions)!.

It deserves mentioning that in contrast with the situation in the Wilson
 theory nothing special happens to the quantities of interest when we
 go through the (2-dimensional!) critical point. As we saw in the
 examples for finite $N$ the area law for the average of a  FWL is equally
 valid both above and below the phase transition: the averages in the  KMM
are related not to correlators of the 2d system,
but to extensive quantities (like density of free energy).

It remains to comment on the very notion of ``temperature''
for these two dimensional systems. As is already clear from the example
of the Ising
model in Section 5, this ``temperature'' is nothing but some function of
the ``frozen''
$\phi$-fields. In the case of $N=2$ it is simply related to the
Boltzmann weights
$C_{ij}$ by Eq.(\ref{(5.2)}). Denominators in this formula
account for
 the normalization condition (\ref{(3'.7)}) for the Boltzmann weights,
which
 is peculiar for the KMM (and in fact guarantees that $\alpha' = 0$
for $T=0$).
This condition cannot however be taken into account in the
same simple way
when $N \neq 2$ since in the latter case the ``bond energies" $J_{ij}$
are $T$-dependent.
One can deal
 with this by assuming that the bond energies also depend on some other
external parameters
(like magnetic fields) and the KMM is associated with some particular
 line in the space of all of these parameters, labeled by $T$.
This remark is intended to alleviate confusion resulting from the
counter--intuitive properties of the system which arise from the
normalization of the Boltzman weights.  The constraints
(\ref{(3'.7)}) guarantee that they are
conditional probabilities which is unusual for Gibbs-like distributions.


Thus we see, that much remains to be done in order to study the continuum
limit of the
KMM, both for finite $N$ (where it may also exist!) and for infinite $N$.
In this paper
 we have tried to show that this study is very much in line with modern
mathematical physics and it should deal with  familiar 2--dimensional
 systems,
their critical behaviour (i.e. conformal models and their perturbations)
 and string--like models.

\section{Conclusion}

In our preliminary investigation of gauge invariant observables in the KMM
(\ref{(1.1)}) we find the model to be very interesting on its own
(irrespective of whether it is related to  4--dimensional
gluodynamics). It is related in a non-trivial fashion to various
statistical systems and a deeper understanding of these connections
seems desirable. Moreover the study of the KMM for finite
values of $N$ seems to be already giving us non-trivial information about
 the
probable large $N$ behaviour -- a  fact which is rather common in  modern
approaches to matrix models.

 As to the relation of the KMM to gluodynamics we  emphasize primarily the
drastic difference between the behaviour of the KMM as the continuum
limit is approached  and that of the conventional Wilson lattice QCD. The
KMM approaches
this limit in a much smoother way. This may be an advantage
of the model, at least from the viewpoint of exact solvability.
It however implies that the transition to the continuum limit is far less
naive than is usually believed. (We illustrated this point with an example of
 the normalization of observable operators.) This phenomenon is
related to the relevance of the angular variables
(rather than only the eigenvalues) of the $\Phi$ field
 in the KMM and it can thus be of  general importance
for all  $c\ge 1$ models.

Another difference between the KMM and the Wilson theory --
the occurrence of extra  $Z(N)$ gauge symmetry --
may  somehow be related to the previous difference. In fact
whenever this $Z(N)$ symmetry is unbroken it comes
 very close to guaranteeing the existence
of a confinement phase all by itself since
all operators with any color properties, even locally,  are simply
not gauge invariant. The operators which can be considered look like
nets of double  lines imitating  planar--diagram
structures.
Despite this essentially confinement--like description, the area law
in the continuum limit seems to arise in a somewhat non-trivial manner
 which
accounts for the actual dynamics of the theory.

We argued that the dynamical behaviour of the KMM can be studied in terms of
{\it 2--dimensional} systems. This fact reflects the already mentioned
 difference between the  KMM (and, in fact, the ``intermediate'' model
 (\ref{(1'.3)})) and the Wilson theory which makes KMM especially
 attractive, namely that the string--like description is essentially built
into
the theory. We have already seen this in several places.
Elementary gauge-invariant observables like the FWL depend  on
a $surface$ rather than just on a contour. Moreover, in reasonable
approximations (like that of the master-field approximation for $\phi$)
their averages
can be evaluated within the framework of auxiliary $2-dimensional$ models.

The operators related to the FWL
 which are presumably the physically relevant ones were  suggested in
(\ref{(2.11)}) to be of the form
$\hat{\cal W}(\Gamma) = \int [{\cal D}S] {\cal W}(S)$
and thus to include sums over all surfaces
with a given boundary. Note, that the averaging procedure, $\ll ~~\gg$,
prescribed by  (\ref{(1.1)}) does not involve any summation of this kind.
It comes entirely from the definition of the observables. Conversely
the ``stringy"
 measure $[dS]$ need not involve the usual $e^{-Area}$ factor since it
appears from the
area law behaviour of $\ll {\cal W}(S) \gg$.

If we now consider the correlator of two Wilson loops the quantity of
interest  is
\bq
G=~\ll \hat{\cal W}(\Gamma_1;\Gamma_2) \gg ~ - ~
\ll \hat{\cal W}(\Gamma_1) \gg ~\ll \hat{\cal W}(\Gamma_2) \gg ~ =
\nonumber
\eq
\bq
\int_{\partial S = \Gamma_1 + \Gamma_2} [{\cal D}S] \ll{\cal W}(S)
\gg ~-~
\eq
\bq
\int_{\partial S_1 = \Gamma_1} [{\cal D}S_1]
\int_{\partial S_2 = \Gamma_2}[{\cal D}S_2]
\ll {\cal W}(S_1) \gg ~\ll {\cal W}(S_2) \gg. \nonumber
\label{(8'.1)}
\eq
The first term  on the r.h.s. contains sums over connected and
disconnected
surfaces (Fig.5). and
\bq
G=~\ll \hat{\cal W}(\Gamma_1;\Gamma_2) \gg_{\rm connected} +
\nonumber
\eq
\bq
\ll \hat{\cal W}(\Gamma_1) \hat{\cal W}(\Gamma_2) \gg ~ - ~
\ll \hat{\cal W}(\Gamma_1) \gg ~\ll \hat{\cal W}(\Gamma_2) \gg
\eq
The last two terms on the r.h.s. are equal to
\bq
\int [{\cal D}S_1][{\cal D}S_2] \left(
 \ll {\cal W}(S_1) {\cal W}(S_2) \gg ~ - ~
\ll {\cal W}(S_1) \gg ~\ll {\cal W}(S_2) \gg \right).
\label{(8'.2)}
\eq
The expression in the brackets obviously vanishes in the master--field
approximation for $\phi$ unless the
surfaces $S_1$ and $S_2$ are
intersecting. However as soon as $D>4$ two 2--dimensional
surfaces in  general locations do not intersect at all. For $D=4$ they
intersect only
at isolated points. It follows that disconnected surfaces do not contribute
to the correlators (perhaps, up to contact terms),
which are thus represented as sums over connected 2-dimensional surfaces
with  a fixed
 boundary distributed according to the area law. (At large
distances the minimal--area surfaces will dominate.)

Another example of the same type arises when the KMM theory is considered
at finite
temperature $\beta^{-1}$ (not to be confused with the ``temperature''
 in our
analysis of auxiliary 2-dimensional systems). What is peculiar to this
situation
in Yang-Mills theory is the occurrence of uncontractable  closed contours
 $C$ which
wrap
around the periodic imaginary time $\tau$ with a period $\beta$.
This contour    has an
 associated Polyakov loop (PL) operator:
\bq
L(C) \equiv W(C)~\stackrel{ncl}{\rightarrow}~~
 {\rm tr}P\exp i\oint_0^{\beta}A_0d\tau
\eq
Usually this operator is the relevant order parameter for the
confinement--deconfinement phase  transition.  It is however
not suitable for KMM since it is not invariant under the
$Z(N)$ gauge symmetry so that $\ll L \gg = 0$ for all temperatures.
This is,
of course, contrary to the usual case where in the high-temperature phase
 $\ll L \gg \neq 0$, signaling the breakdown of a $global$
  $Z_N$-symmetry.
Our standard trick of ``filling'' the Polyakov loop is not applicable
since the contour in uncontractable.
What can be considered instead in the KMM is the counterpart
of a correlator of $two$ Polyakov loops (Fig.6):
\bq
\ll{\cal L} (C , C')\gg~ \equiv L(C)L(C')\prod_{\Box \in S(C,C')}W(\Box)
\eq
One can consider the behaviour of this ``Filled Polyakov Loop" (FPL)
when the distance  $R$ between $C$ and $C'$ is large.
In analogy with Polyakov's result for the ordinary QCD \cite{P}
one can suggest that there are two possible regimes:
\bq
\ll{\cal L(C,C')}\gg \rightarrow e^{-m(\beta) R};~~ \beta >
 \beta_{\rm crit}
\nonumber \\
\ll{\cal L(C,C')}\gg \rightarrow  const;~~ \beta <\beta_{\rm crit}
\eq
and there is a confinement deconfinement transition at some critical
inverse temperature $\beta_{\rm crit}$,
such that $m(\beta_{\rm crit}) = 0$.
\footnote{ What is  amusing is that precisely the same configuration
 was suggested in \cite{BK} to describe  the vortex-antivortex
 configuration in the matrix model for the $d=1$ string. One can
consider the boundary  Polyakov lines $L(C)$ and $L^{\dagger}(C')$
as a vortex and antivortex on the world sheet given by the net
 $\prod_{\Box\in S(C,C')} W(\Box)$. This  coincidence is not accidental
and is connected to the fact that the confinement--deconfinement
 (or Hagedorn ) transition  is equivalent to the
 Berezinski-Kosterlitz-Thouless (BKT) transition on the world-sheet
 \cite{KOG} which involves condensation of the vortex-antivortex
 pairs. The critical phase transition occurs when the vortex-antivortex
configurations become unsuppressed,
i.e. precisely when  $\ll{\cal L(C,C')}\gg ~\neq 0$. Analogy between
the confinement-deconfinement transition
in large $N$ QCD and the BKT transition on the world sheet was also
considered in \cite{AW}.}

 Thus, if one accepts that observables in KMM can be represented as sums
over fluctuating $2d$ surfaces, then their contractable boundaries
(Wilson Loops) play the role of the physical states (gluonia), while
uncontractable boundaries (Polyakov Lines) describe  (world-sheet)
vortices.
We see, that in contrast to the Wilson theory the string--like
description is somehow built into the KMM. It just cannot be treated
in any other terms. This may be the main reason why this model
should be studied in more detail and why one may ultimately
 believe in its relevance for the theoretical analysis of realistic QCD.

\vskip 2.0in
{\bf Acknowledgments}

	 We are indebted to Yu.Makeenko for discussion of $Z_N$-invariance
and to A.Niemi and S.Shatashvili for that of unitary group integrals. A.M.
acknowledges the hospitality of the theory group of the Physics Department
   at  UBC.

\newpage
\vskip .3in
\bf{Figure Captions:}
\vskip .3in

\noindent
Fig. 1 ~~Adjoint Wilson Loop for a rectangular contour $\Gamma$.

\medskip
\noindent
Fig. 2  ~~ Filled Wilson Loop for a rectangular surface $S$.

\medskip
\noindent
Fig. 3 ~~ Example of the substitution of the ``fat graph''
$\Gamma_1+\Gamma_2+...\Gamma_6=0$ by an ordinary graph $\tilde\Gamma$
according to the identity (\ref{(4.1)}).

\medskip
\noindent
Fig. 4 ~~ Examples of contours with ``inverse steps'' which are
included in the sums (\ref{(1.4)}) and (\ref{(6".13)}). Only contours
of the type (b) exist in the case when the lattice is a one
dimensional chain.

\medskip
\noindent
Fig. 5 ~~ The identity $\ll\tilde{\cal W}(\Gamma_1;\Gamma_2)\gg
= \ll\tilde{\cal W}(\Gamma_1,\Gamma_2)\gg_{\rm connected}+
\ll\tilde{\cal W}(\Gamma_1)\tilde{\cal W}(\Gamma_2)\gg$.

\medskip
\noindent
Fig.6 ~~ The surface $S$ arising in the definition of the
filled Polyakov Line $\cal L(C,C')$.

\end{document}